# What is the Price of a Skill?
# The Value of Complementarity

September 2023

Fabian Stephany ◆ *         Ole Teutloff♥*

## Abstract

The global workforce is urged to constantly reskill, as technological change favours particular new skills while making others redundant. But which skills are a good investment for workers and firms? As skills are seldomly applied in isolation, we propose that complementarity strongly determines a skill's economic value. For 962 skills, we demonstrate that their value is strongly determined by complementarity – that is, how many different skills, ideally of high value, a competency can be combined with. We show that the value of a skill is relative, as it depends on the skill background of the worker. For most skills, their value is highest when used in combination with skills of a different type. We put our model to the test with a set of skills related to Artificial Intelligence (AI). We find that AI skills are particularly valuable – increasing worker wages by 21% on average – because of their strong complementarities and their rising demand in recent years. The model and metrics of our work can inform the policy and practice of digital re-skilling to reduce labour market mismatches. In cooperation with data and education providers, researchers and policy makers should consider using this blueprint to provide learners with personalised skill recommendations that complement their existing capacities and fit their occupational background.

**Highlights:**

- We expand the literature on human capital formation by quantifying the market value of skills.
- We demonstrate that the value of a skill depends on complementarity, that is, by the number, diversity, and value of skills it can be combined with.
- We reveal that the value of a skill is relative; it depends on the type of skills it is combined with.
- We showcase that AI skills are particularly valuable as they have high levels of skill complementarity, increasing worker wages by 21% on average.

◆ Oxford Internet Institute, University of Oxford, Oxford; Humboldt Institute for Internet and Society, Berlin; Bruegel, Brussels, Belgium;  fabian.stephany@oii.ox.ac.uk

♥ University of Copenhagen, Centre for Social Data Science * *Both authors contributed equally to the work.*

# Introduction

Will we be racing against or with the machines? (Brynjolfsson & McAfee, 2012). Today, our prospects as workers and firms depend on how well we can understand and adapt to technological change, with one of the defining features of this change being its relationship to our skills, given that technology is not "skill neutral". While some jobs might indeed disappear due to technological change, the remaining ones will change, and entirely new jobs will emerge (Acemoglu & Autor, 2011; Brynjolfsson & Mitchell, 2017; Frey & Osborne, 2017). Similarly, technologies, such as Artificial Intelligence (AI), create new occupations characterised by new tasks that require novel sets of skills. Because the work that is eliminated has different skill requirements than the newly created jobs, this results in the paradoxical situation of simultaneous unemployment and labour shortage (Autor, 2015). In other words, workers risk being pushed out of employment at the same time as companies struggle to find suitable employees to pursue new types of jobs. To stay in employment, workers therefore need to learn new skills and combine them with existing skills in novel ways. To stay competitive, employers need to invest in reskilling their workforce and talent acquisition. However, for many of the newly emerging jobs, precise skill requirements are unclear and constantly evolving.

Against this background of widespread uncertainty on how to address the growing skill mismatch and resulting labour market inefficiencies, the conventional policy response – aligning training programmes with changing labour market demand – is becoming increasingly ineffectual as technological and social transformation outpaces national training systems (Collins & Halverson, 2018). Likewise, large employers struggle to keep the skills of their workforce up to date (Illanes et al., 2018). Employers, workers and education providers seem uncertain about which new, often digital, skill is the first step towards a successful re-skilling trajectory – should workers be learning to work with AI, and if so, should they be concentrating on programming in Python[1] or Java[2]? Or on something else entirely? This challenge is particularly complex for workers, who have already spent considerable time in the labour market and do not have the resources to build their skill portfolio anew from scratch. They are trying to find synergies between their existing skill set and new capabilities. In this setting, policy makers, businesses, and workers are striving to find skills that ensure them a sustainable position in the future workplace. But how can they decide which skills to invest in?

---

[1] Python is a general purpose programming language of immense popularity. It caters to a vast spectrum of domains, including but not limited to data science, software and web development, automation, and overall task execution optimization.

[2] Java, another very popular programming language, distinguishes itself as a platform-independent framework. It is frequently used for running mobile applications, building and scaling cloud applications, and developing chatbots and other marketing tools.

In this work, we argue that complementarity is essential for estimating a skill's value. There are, at least three reasons, why complementarity matters. For one, skills come in sets. The value of a skill can't be assessed in isolation, as it is rarely applied in isolation – our dataset alone counts on average 80 skills per occupation. Furthermore, for reskilling, complementarity is a fundamental feature, as we do not build our skill set from scratch but rather, incrementally, add new skills to existing capacities. In this process, economic efficiency is relevant, as we, ideally, want to make use of as many complementarities between old and new skills as possible. Lastly, we assume that a skill's complementarity reflects its strategic value. The more diverse the set of capacities a skill can be combined with, the more options a worker has for reskilling – increasing their resilience against unforeseen technological changes in the future.

We test these assumptions and show that the value of a skill depends on its complements in at least three ways. First, complementarity means options: A skill is more valuable, if it can (potentially) be combined with many other skills. Secondly, we show that not only the number but also the value of complements matters: A skill is more valuable, if its complements are of high value, too. Thirdly, we show that the value of a skill is relative, as one and the same type of skill can have different premia, if combined with a different set of skills that a worker already has. In addition to complementarity, we show that skills are more valuable if they enjoy high levels of market demand relative to worker supply. We put our model to the test with a set of skills used to develop and maintain AI technologies. AI skills, such as programming languages and data analytics, are particularly valuable, as they have strong complementarities in terms of number, diversity, and value of complements, as well as a high level of demand.

Our paper adds to four related streams of literature. Most importantly, we contribute to an established debate on how to measure human capital. We attach an interpretable market value to individual skills and expand the theory on skill complementarity. Secondly, we build on existing attempts to model the relatedness between skills using network science. We derive communities of skills from this network approach and describe the value of each skill by the characteristics of its closest neighbours. To guide our empirical analysis, we synthesise these scholarly debates into hypotheses about what determines the value of a skill. Thirdly, we contribute to scholarly work on how technological change and automation shift the demand for skills. Here, our work suggests a pragmatic approach to the near real-time monitoring of demand and value of individual skills in a fast-changing labour market. Lastly, we contribute to the growing body of research that uses digital trace data to study labour market developments. In contrast to recently suggested data sources that examine either the market's supply or demand side exclusively, our approach takes both demand and supply into account and, crucially, allows us to attach price tags to individual skills.

Our findings are applicable in several ways. They allow us to identify and track over time what skills and what *combinations* of skills are in demand and successful (as an economic premium). Conceptualising the relationship between skills as a network enables us to identify potential individual learning trajectories, finding the most valuable complement to add to any worker's existing skill bundle. Thereby, we could support workers in building individually tailored, data-driven reskilling pathways. We illustrate our method for skill evaluation with online freelancing data for U.S.-based workers, however, the method can be extended to other data sources, such as online job ads or profiles from career building webpages, that cover larger parts of the traditional labour market.

# Background

## The Economics of Skills

The periodic warning that automation and new technologies are going to replace large numbers of jobs is a recurring theme in economic literature, most recently in the case of digital and AI technologies (Acemoglu & Autor, 2011; Brynjolfsson & McAfee, 2014; Frey & Osborne, 2017). However, measuring the effects of new technologies on the future of work is not trivial. Looking at education and wages at the aggregate level alone has proven to be insufficient (Acemoglu & Restrepo, 2018; Beaudry et al., 2016; Brynjolfsson & McAfee, 2014), and treating occupations as homogeneous entities with a certain automation probability risks being overly simplistic and could even lead to false conclusions (Frank et al., 2019). Skill requirements of occupations are dynamic, because technological innovations change the demand for specific skills and thereby the skill composition of occupations – a phenomenon known as skill-biased technological change (SBTC) (Acemoglu & Autor, 2011).

Before skills, researchers initially examined the selective effects of technology on tasks (Autor et al., 2003) to account for the correlation between the usage of computers (or other automating technologies) and the automation of specific job task content. The task-based approach emphasises how technology can supplement certain tasks while substituting for others. To understand how this selective automation of tasks translates into changing skill requirements, scholars established a relation between tasks and skills (Acemoglu & Autor, 2011; Autor & Handel, 2013): Tasks refer to the features of a job (demand side), whereas skills describe the attributes of workers (supply side). Workers acquire skills that give them an edge in particular tasks. As technology automates specific tasks and hence the demand for the respective skills – rather than entire occupations – economists suggest that it's best to understand occupations as "bundles of skills".

Placing skills at the core of investigating technological labour market change has opened an entirely new field of economics of skills (Nedelkoska et al., 2019). In the context of SBTC but also for judging the effects of outsourcing and offshoring on the labour market (A. Blinder, 2007; A. S. Blinder & Krueger, 2013), economists initially distinguished among skills for solving routine, non-routine, and interactive job tasks on the one side and between manual and cognitive tasks on the other side, suggesting that computers augment non-routine and interactive tasks, but substitute for routine manual and routine cognitive tasks (Autor et al., 2008; Goos et al., 2009). The SBTC hypothesis was further developed by Acemoglu & Autor (2011) , who proposed that computers constitute the primary demand-side technology, negatively affecting only routine tasks. These routine tasks were viewed as the domain where medium skill workers held a comparative advantage. However, there is now growing scepticism about all of these assumptions, as scholars contend that AI and robotics are able to substitute for non-routine tasks, both cognitive and manual (Brynjolfsson & McAfee, 2014; Frey & Osborne, 2017).

The economics of skills helps us to gain a better understanding of technological impacts on work, as it allows us to investigate the relationships between skills and their potential economic synergies. The influence of SBTC but also the growth of the knowledge economy (Powell & Snellman, 2004) have intensified the need for a more sophisticated understanding of skills and their relatedness (Aggarwal & Woolley, 2013; Ren & Argote, 2011). Previous perceptions of viewing skills as interchangeable oversimplified the complexity of human capital, as workers possess a variety of heterogeneous and multidimensional skills that match the demands of different jobs (Cunha et al., 2010). A significant amount of research has demonstrated the critical role of skill diversity, specialisation, and recombination in problem-solving and knowledge creation (Hong & Page, 2004; Lazear, 2004; Woolley et al., 2010).

## *The Complementarity of Skills*

A central theme in the economics of skills is the idea that economic returns on skills are not only affected by different individual skills but also by the "balancing" of these skills (Lazear, 2004), leading to the question of whether workers should become generalists ("balanced skill set") or specialist ("unbalanced skill set"). While there is no definite consensus on this trade-off between breadth and depth, research underscores that specialists often tend to have the upper hand (Buchen et al., 2021; Leahey, 2007; Waller & Anderson, 2019), while the context of work roles sometimes favours a generalist strategy, e.g., for entrepreneurs (Bublitz & Noseleit, 2014). The underlying assumption of specialisation is synergies arising from skill complementarities (Allinson & Hayes, 1996). For some skills (e.g., logo design and illustration work) it is fairly obvious that skill complementarities will emerge. The bundle of logo design and illustration work is more valuable than the sum of its parts, given that jobs in graphic design, brand or product design, for example, would typically require both. Oftentimes, skill complementarities are likely to be limited to their respective occupational domains, e.g., logo design and translating Korean should have little complementarity, while being valuable skills in other contexts (Anderson, 2017; Stephany, 2021). Consequently, the value of an extra skill is contingent upon the worker's existing skillset (Agrawal et al., 2015; Altonji, 2010; Autor & Handel, 2013).

Oftentimes, when assessing the economic returns of skills, e.g. in forms of earning premia, scholars distinguish between cognitive and noncognitive skills.[3] There is general consensus in the literature that noncognitive skills, in general, are associated with a positive return in the form of better employability, higher earnings, or increased wellbeing (Cabus et al., 2021; Greiff et al., 2015; Heckman et al., 2006). However, some noncognitive skills, such as neuroticism (i.e. low emotional stability) have been shown to reduce economic returns (Cabus et al., 2021). With the advent of digital automation, the economic return from noncognitive skills has increased. Edin et al. (2022) show that workers with strong noncognitive skills are increasingly moving to occupations that require abstract, non-routine, social, non-automatable and non-offshorable tasks, which offer higher wage premia. But noncognitive skills also matter in combination with cognitive skills – Palczyńska (2021) shows that the positive returns on cognitive skills are higher among more non-neurotic individuals. For digital sectors, the complementarity or bundling of cognitive and noncognitive skills is particularly relevant. In a comparison of 31 countries on adult skills and labour market participation Grundke et al. (2018) show that within digital-intensive sectors, there is a distinct preference for employees who possess elevated levels of advanced numeracy skills in combination with self-organisation, effective management, and communication skills.[4]

---

[3] In education studies, cognitive skills are described as competencies acquired through education such as literacy and numeracy, or specific technical skills like coding or statistics, while non-cognitive skills are used as an umbrella term for characteristics such as motivation, social skills, perseverance, etc. (Demange et al., 2022).

## *The Relatedness of Skills*

The availability of novel data sources representing manifold aspects of relatedness (Hidalgo, 2021) has enabled scientists, most prominently in the field of complexity economics, to capture the relationship and synergies of skills via network models. Waters & Shutters (2022) examine a U.S. skill network, in which skills are connected if performed by the same worker, focusing on the relationship between network centrality of skills and economic performance. They find that occupations with higher skill centrality are associated with greater annual salaries, and metropolitan areas with higher skill centrality have higher productivity rates. These results suggest that the application of traditional network metrics to this view of cities as complex systems can offer new insights into the dynamics of regional economies. When it comes to the complementary value of a skill, Dave et al. (2018) show that the relative positioning of a skill in a skill network contains information on how likely it is for a person to acquire it. In their simulation of skill acquisition they show that a skill is more likely to be acquired by learners if it is close to the bundle of their already existing capacities.

---

[4] Our data do not allow us to explicitly measure the effect of noncognitive skills in our analysis. However, we try to account for the variability in noncognitive capabilities, such as neuroticism or work attitudes, as we control for characteristics on the worker level when estimating the value of cognitive skills.

Similarly, del Rio-Chanona et al. (2021) use network metrics to track worker transitions between occupations due to automation shocks. They show that transitions between occupations are more likely if skill similarities are high. This leads to the assumption that certain "bridging" skills are particularly valuable, as they enable a transition from one (less profitable) occupation to the next. This complementary aspect of skills is outlined by Alabdulkareem et al. (2018) who argue that much of the polarisation across industries and occupations can be described by skill polarisation, which describes groups of workers that are "stuck" in a low-value segment of the skill network with little proximity to the central skills that would allow them to shift into more profitable occupations.

Recognizing the importance of skill complementarities, Neffke (2019) investigates the value of having complementary co-workers using a large Swedish dataset on educational specialisations. Introducing educational synergy and "educational substitutability" allows Neffke (2019) to measure and model co-worker complementarity, showing that the value of what a person knows is relative, as it depends on the skillsets of those with whom they work (Neffke, 2019). The work shows that cross-worker complementarities, emerging from the interaction of a set of specialised workers, e.g., in R&D or product design, are highly relevant.

While the above contributions highlight the importance of understanding the relatedness of skills, they do not explicitly seek to measure and describe the value and complementarity of skills. Our contribution is to return to the assumption that a skill's value is, in part, determined by the number, diversity, and value of its complements. Modelling the relationship between skills as a network allows us to map skills by their complementarity, with skills that sit closely in the network having higher levels of complementarity. Accordingly, our work focuses on the complementarity between individual skills. We model this complex relationship using network science, and confirm the importance of complementarity in three ways. First, we find that the economic value of a skill increases with the number of diverse complements that it can have – this is irrespective of the absolute popularity of a skill, e.g., how often it is demanded on the market. Secondly, we find that the value of a skill is positively related to the value of its complements. Thirdly, we find complementarity between skills and workers. That is, skills that allow workers to diversify their skill portfolio are of higher value to them than skills that are close to their existing skill domain. This illustrates the relative value of a skill, as it depends on a worker's existing skill bundle.

## What Determines the Value of a Skill?

Our work aims to describe the market value of a skill. Based on the theoretical considerations outlined in the previous sections, we argue that complementarity influences the market value of a skill. In addition, we assume that skill premia are higher for capacities with high levels of labour demand and low levels of labour supply, as we summarise in the following:

1) **Hypothesis 1 – Market.** Our assumption rests on an intuitive understanding of market dynamics. We perceive the value of a skill as the premium it commands, which is conventionally determined by forces of supply and demand. ***We postulate that a skill that is often requested and seldomly offered by workers is of high value.***

2) **Hypotheses 2 – Complementarity.** In accordance with the literature on skill relatedness and specialisation we further assume that the value of a skill depends on its complementarity. Literature on the relatedness of skills (Nedelkoska et al., 2019) suggests that skills have different degrees of recombination, that is, options to combine them with other skills into marketable bundles. In a setting where the capability to reskill determines workers' competitiveness, the degree of recombination contributes to a skill's value. ***Accordingly, we postulate that the value of a skill is higher if it can be combined with a diverse set of other skills of high value.*** This complementarity is reflected by at least three aspects:

    a) **Number of Complements.** The number of adjacent skills should be positively related to a skill's value.

    b) **Diversity of Complements.** The diversity of adjacent skills should be positively related to a skill's value.

    c) **Value of Complements.** The value of the adjacent skills should be positively related to a skill's value.

# Method

## *Online Labour Platform Data*

The data for this analysis stems from one of the most popular online freelancing platforms[5], also referred to as online labour markets as introduced by Horton (2010). These platforms are websites that mediate between buyers and sellers of remotely deliverable cognitive work (Horton, 2010). The sellers of such work are either people in regular employment earning additional income by moonlighting via the Internet as freelancers, or they are self-employed independent contractors (Stephany et al., 2020a). The buyers of work range from individuals and early-stage startups to Fortune 500 companies (Lehdonvirta & Corporaal, 2017). Online labour markets can be subdivided into microtask platforms, e.g, Amazon Mechanical Turk, where payment is on a piece rate basis, and freelancing platforms, such as Fiverr, where payment is on an hourly or milestone basis. Between 2017 and 2020, the global market for online labour grew by approximately 50% (Stephany et al., 2021). Online labour market data allow us to monitor skills in a global workforce on a granular level and in near real-time. The data include worker and employer location, project wages, previous income, and project-level skill requirements.

The platform of our analysis is one of the most popular online freelancing places globally. Most recent estimates state that it records more than 180,000 active workers globally in 2021 (Kässi et al., 2021). The number of projects in our dataset are evenly distributed along the observation period (2014–2022) with roughly between 6000 and 7000 projects a year completed by online workers located in the United States. Similar to the overall development of the online freelancing market (Stephany et al., 2020b), the number of projects decayed during the first months (March to May 2020) of the Covid-19 but quickly bounced back to pre-pandemic levels in late 2020. While in general a few online labour market projects remain unfilled, our sample only includes completed projects, of which 85 percent lasted between one and seven days on average. As is usually the case for this type of online freelancing platform, all projects were undertaken remotely and workers completed their projects individually without cooperating with each other. While both hourly and flat rate projects are listed, we only consider hourly jobs in calculating the wage rate, because we do not observe the hours worked on fixed-price jobs. The average worker on the site makes \$25/h (median) and lists six to seven skills.

---

[5] The platform wished not to be identified by name (for details see Lehdonvirta et al., 2019).

Freelancing plays a significant role in the U.S. labour market. According to the UpWork Freelance Forward 2022 survey (Upwork, 2022), freelancing contributed about \$1.35 trillion to the U.S. economy in 2022 with an estimated 60 million Americans engaging in some form of freelancing. About 31 million of them provided knowledge-based services like programming, marketing, IT and business consulting. Freelancing is growing notably among highly educated workers. An estimated 26% of U.S. freelancers hold postgraduate degrees. In particular, online freelancing, that is completing work via online platforms, has become a phenomenon of recognisable size in the U.S., with the platform providing data for our investigation being the country's most popular marketplace for online freelancing. It is difficult for us to show that the workers in our dataset are representative of the overall U.S. freelancing population, as the freelance population is very heterogeneous, difficult to measure and we lack important demographics of the online freelancers in our dataset, e.g., age, gender, and formal education.

However, establishing representativeness, in terms of worker characteristics and working conditions, is not the goal of our analysis. The focus of our analysis is on how workers use and combine skills. In traditional labour market settings (regular employment and freelancing) we typically can't observe workers' skills, and we propose online freelancing data as an innovative data source to address this gap. Since we can't observe skills in traditional labour markets, we can't know if the way online freelancers use skills is representative of the wider workforce. It seems plausible that there might be differences with regards to soft skills and other noncognitive skills. With regards to hard skills, differences should be smaller; that is, designers, programmers or writers require similar hard skills whether they find their work via an online platform or not. Our analysis focuses on this type of hard skills. While online freelancing data is certainly context-specific, we argue that the fundamental dynamics of demand, supply, and complementarity that drive the value of skills should not be drastically different in other segments of the labour market.

*Building a Network of Skills*

We conceptualise the relationship between skills using a network approach. The data consists of a sample of 49,884 freelance projects posted between 2014 and 2022 with multidimensional skill requirements completed by 25,170 U.S.-based workers. We constrain our analysis to U.S.-based workers to limit variation in wages driven by structural differences between countries (e.g. local price levels) and other unobserved heterogeneity. We define a skill as a capacity that a worker needs to perform certain job tasks (Nedelkoska et al., 2019). However, we acknowledge that skills are typically more general than tasks: programming skills allow a person to perform a range of computational and analytical tasks, and writing skills can be applied in numerous job tasks related to producing text creatively. In this regard, the set of skills considered in our case covers three levels of hierarchy. Some skills are used to complete one specific task, such as *photo retouching*. Other skills can be used for solving various tasks, such as working with photo editing software, like *Adobe Photoshop*. Lastly, some skills cover an entire domain of tasks, such as *photo editing*. At first, these different levels might suggest redundancies in the description of skill portfolios, e.g., skills in *photo retouching* naturally require the capacity to work with software like *Adobe Photoshop* and know-how in *photo editing*. However, the opposite is not the case; skills in *Adobe Photoshop* extend far beyond *photo retouching* alone and skills in *photo editing* do not necessarily require skills in *Adobe Photoshop* or *photo retouching*.

Using the information on skill requirements contained in each project, we construct a network in which 962 skills are represented as nodes that are connected by a link if a worker applies both in a freelance project.[6] The links are weighted according to how often two skills co-occur, as illustrated in Figure 1. The more often they co-occur the more related they are. Thereby, we build on the established concept of relatedness of economic activities (Hidalgo, 2021) to conceptualise the relationship between skills. Using the method of skill networks and relatedness allows us to move our analysis beyond the pairwise likelihood of two skills appearing together. Instead of simply measuring how often two skills occur in the same project, we can answer questions about how skills relate to each other even if they are never jointly applied. For example, the network structure determines that *logo design* and *creative drafting* are quite closely related. Even if they

---

have never been jointly applied in a single project, they are embedded in a broader "neighbourhood" of skills and linked by intermediaries, such as specific software packages like *Adobe Illustrator.*[7] To categorise skills in distinct groups or "neighbourhoods" we use community detection, one of the foundational methods in network science. This method has been successfully applied to data on skills in various contexts illustrating for example how O*NET skills cluster into socio-cognitive skills and sensory-physical skills (Alabdulkareem et al., 2018; Water & Shutters, 2022), how freelancer skills form human capital networks (Anderson, 2017) or how workers with different educations connect to create synergies (Neffke, 2019). Speaking more formally, the skill network provides us with an endogenous categorisation of skills and illustrates the context dependency of human capital.

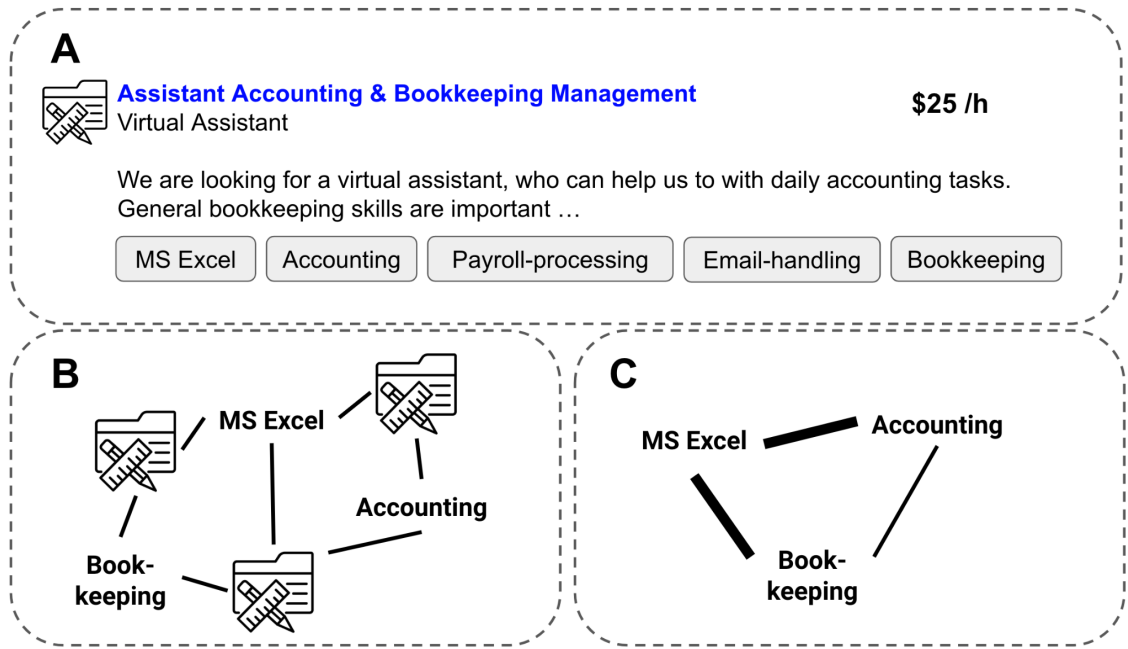

**Figure 1:** *Derivation of a unipartite network of skills (C) from the bipartite network connecting freelancer projects and skills (B), where two skills are connected if a worker applies both in a particular freelance project (A).*

## *Measuring the Value of Skills*

The hourly wage in USD associated with each project allows us to statistically assess the value of individual skills. We calculate this *premium* of skills using a regression approach proposed by Alekseeva et al. (2021). In our case, 962 linear regression models – one for each skill – are used to explain a project's hourly rate with:

---

[7] Adobe Illustrator is a software application for creating drawings, illustrations, and artwork.

1. the year the project has been carried out in (yearly dummies, 2014–2022),

2. the occupational category the project falls into (12 categories[8]),

3. the experience of the worker (number of past projects),

4. the occurrence of a specific skill (skill dummy).

Each of the 962 most popular skills is considered individually as an explanatory feature in the linear regression. The beta coefficients of each of the 962 skill regressions allow us to determine the added value of an individual skill according to the following formula:

$$log(project\,wage_k) \,=\, \beta_0 + \beta_1 * year_k \,+\, \beta_2 * occupation_k + \beta_3 * experience_k \,+\, \beta_4 * skill_j \,+\, e_{k'}$$
$$k \in 1,...,n\ and\ j \in 1,...,962$$

As the dependent variable, the project wage in USD per hour, is log-transformed, we therefore read the coefficient of the skill dummy $\beta_4$ as percentage change in wages. Performing this model 962 times for each skill provides us with added values, called premia, for each skill.

Most projects require more than one skill. In this case, the project is counted for each skill it requires. By calculating this premium, we derive the market value of the 962 most popular skills, that is skills that occur in at least 20 projects.[9] We include additional controls, as it is likely that factors other than the presence of a specific skill determine project wages. For example, it could be imagined that annual price levels, the location of a worker, the duration of a project, or her working attitude, e.g. friendliness or punctuality, etc., exogenously influence the project wage. We try to capture these latent influences, as we can control for various characteristics, such as the type of job, the worker experience and the year that it has been carried out (the geography of all projects is limited to the United States). To test the sensitivity of project wages to latent characteristics we calculate the so-called skill ratio in addition to the skill premium. The ratio is simply the relative increase in hourly wages of projects that contain a specific skill to those that do not (see Section S2 and Figure A3 in the Appendix). Without any controls in the wage regression, ratio and premium are almost perfectly correlated. Once including controls on the worker and project level, this (Pearson) correlation is reduced, indicating that both metrics consistently measure a similar concept, while limiting the influence of exogenous factors.

---

[8] We use the following occupational categories provided by the platform: "Sales & Marketing", "Web, Mobile & Software Dev", "Writing", "Design & Creative", "Translation", "IT & Networking", "Data Science & Analytics", "Admin Support", "Customer Service", "Legal", "Accounting & Consulting", "Engineering & Architecture".

[9] This popularity threshold has been validated empirically, as illustrated in Figure A1 in the Appendix.

Lastly, with our metric of skill premium at hand, we investigate how the variance in values can be described by the features proposed in our hypotheses. We build a regression model that allows us to test our assumptions about what correlates with skill premia:

$$Premium_j = \beta_0 + \beta_1 * Supply_j + \beta_2 * Demand_j + \beta_3 * Degree\ Centrality_j + \beta_4 * Complements'\ Diversity_j + \beta_5 * Complements'\ Value_j + e_j,$$
$$j \in 1,\ldots,962$$

We include the following features:

1. Supply (Number of workers commanding respective skill, log)
2. Demand (Number of projects requesting respective skill, log)
3. Complementarity
   a. Number of adjacent skills (Degree centrality, log)
   b. Diversity of adjacent skills (1 – Gini Coefficient)
   c. Value of adjacent skills (Average value of three most adjacent skills)

For modelling complementarity, we rely on three concepts: the number of complements, their diversity, and their value. The number of complements is represented by the degree centrality of each skill, borrowing this concept from various network science applications. For the examples of aviation markets (Chung et al., 2020) and citation networks (Ding, 2011) it has been shown that the complementarity of nodes, such as relevant airport hubs or popular scholars, can be described by centrality. In addition, the diversity of complements reflects how strongly a skill is related to a specific set of complements. While two skills could have the same number of complements, they might be differently distributed. A skill could very frequently occur together with a core set of other skills and only seldomly with a set of peripheral complements. Or a skill could be regularly recombined with a large set of complements. According to the theory of skill bundling, we assume that skills with a higher diversity in complementarity are more valuable in terms of reskilling, as they allow for a more flexible recombination. We capture the diversity of complements by calculating the Gini coefficient (Sitthiyot & Holasut, 2020) of the frequency distribution of adjacent skills.[10]

Lastly, the value of complements is calculated as the average premium of the three most adjacent skills. However, given the construction of our value metric, some of the correlation in skill premia might be explained by the fact that both a skill and its complements share a joint set of projects. We would like to avoid this property, as it might introduce endogeneity into our model. We achieve this by calculating the complements' premia only based on the projects, which they do not share with the respective skill. Section 2 of the Appendix explains this in more detail and shows that the optimal number of complements to consider for our analysis is three. We take a critical stance on potential endogeneity problems with this model. While we think that complementarity and community are exogenous to the other characteristics, modelling supply and demand usually suffers from issues of multicollinearity. However, we do not observe market supply matching demand in the  particular online labour platform we analyse here, given there are many workers registered without completing any project (Kässi et al., 2021).

# Results

## *Taking a Network Perspective on Skills*

We begin modelling the relatedness of skills by creating a network, as described in the Method section. Figure 2A illustrates the resulting network of skills, i.e the "skill space". We use the Louvain method for community detection (Blondel et al., 2008), with skills clustering into seven communities. We label the clusters based on the type of most prevalent skills in each of the communities: "Finance & Legal, "Software & Tech", "Marketing", "Design", "Audio & Video", "Writing", and "Admin". In addition to the community labels, Figure 2A also highlights three of the most prevalent skills for each of the communities. This network structure allows us to categorise skills but still relate them in terms of their distance within and across categories. For example, it is worth noticing that the skill *Creative Writing* is adjacent to the Design community, while it is more distant to the writing skill *French Translation*. Another example is the Audio & Video skill *Motion Graphics*, which is closer to the Design skill *Drafting* than to the same cluster skill *Music Composition*.

---

[10] We calculate 1–Gini as our diversity measure, since the Gini ranges from 0 (evenly distributed) to 1 (most strongly concentrated). Accordingly, in our calculations, a value of 1 would mean that 20 adjacent skills co-occur in 5% of the projects, while a value of 0 would indicate that only one adjacent skill co-occurs with the respective skill.

The skill communities are different from occupations or industries. They are a bottom-up categorisation of skills, while occupations and industries are a top-down categorisation of workers. To a certain extent these skill communities overlap with the occupation taxonomy provided by the freelance platform (see Figure A2 in the Appendix), i.e., the majority of skills in the Design community also fall into the occupation category Design & Creative. However, other communities, such as Admin, distribute more broadly across the occupations of Writing, Admin & Support, and Legal. This is in line with findings by Anderson (2017) and matches the categorisation of online labour market skills by Kässi & Lehdonvirta (2018). But in contrast to official occupation taxonomy or industry classifications, our clustering allows us to fully group all skills into communities based on their *actual* application and not based on an external categorisation. For example, working with *MS Word* might be defined as an IT skill, while our clustering reveals that it is predominantly applied together with Admin skills. In the next stage of the analysis this will help us to evaluate the contribution of community membership to a skill's market value.

***Figure 2 (A)*** *Skills cluster into seven groups according to their field of application. We label the clusters based on the type of most prevalent skills in each of the communities: "Finance & Legal, "Software & Tech", "Marketing", "Design", "Audio & Video", "Writing", and "Admin". **(B)** Valuable skills – the node size*

*represents the premium of each skill – are not distributed at random across the skill space. We use the software Gephi and the Force Atlas algorithm to layout the network (Bastian et al., 2009).*

## Revealing the Complementary Value of a Skill

Skills have different market values. Table A2 in the Appendix shows the top 20 and bottom 20 skills by premium. While the exact monetary values are specific to one online labour platform in the U.S. they allow us to make quantitative comparisons between skills. We see skills from the domain of Admin work with relatively low skill premia, while skills from Software & Tech or Finance & Legal show higher values. Figure 2A reveals that valuable skills – with the node size representing the value of each skill – are not distributed at random across the skill space. Instead, they seem to have distinct characteristics and a distinct positioning in the skill space. Furthermore, valuable skills are not equally distributed across skill communities, as Figure 2B shows. Skills in the domain of Finance & Legal have, on average, a significantly higher premium than skills in Marketing, which have in return a higher premium than skills in Admin.

To further test our hypotheses, we run multiple regression models on the skill premium, as described in the Method section. In a stepwise fashion, our models include supply (number of projects), demand (number of workers), and complementarity measures of the respective skill. The results of the regression analysis are shown in Table 2.

*Table 2.* The value of a skill, measured by skill premium in models 1-5 is determined by supply, demand, community, and complementarity. The number of complements introduced in model 3 is represented by the degree centrality of each skill and the diversity of complements in model 4. Model 5 introduces the value of complements. In model 5b, we run our regression analysis with a random effects model for skill communities.

| | DV: Premium of a Skill (%) | | | | | |
|---|---|---|---|---|---|---|
| | Model 1 | Model 2 | Model 3 | Model 4 | Model 5 | Model 5b Cluster RE |
| **Market** | | | | | | |
| Supply | -0.29*** | -0.22*** | -0.26*** | -0.28*** | -0.27*** | -0.17*** |
| *Number of worker (log)* | (0.05) | (0.05) | (0.05) | (0.05) | (0.05) | (0.05) |
| Demand | 0.24*** | 0.18*** | 0.20*** | 0.21*** | 0.20*** | 0.15*** |
| *Number of projects (log)* | (0.05) | (0.05) | (0.05) | (0.04) | (0.04) | (0.04) |
| **Cluster** | | | | | | |
| *Reference: "Admin + Mgmt"* | | | | | | |
| Audio & Video | | -0.04 | -0.01 | 0.03 | 0.16*** | RE |
| | | (0.03) | (0.03) | (0.03) | (0.04) | |
| Design | | 0.01 | 0.03 | 0.05** | 0.14*** | RE |
| | | (0.03) | (0.03) | (0.02) | (0.03) | |
| Finance & Legal | | 0.17*** | 0.18*** | 0.18*** | 0.15*** | RE |
| | | (0.03) | (0.03) | (0.03) | (0.03) | |
| Marketing | | 0.10*** | 0.08*** | 0.08*** | 0.06** | RE |
| | | (0.03) | (0.03) | (0.02) | (0.02) | |
| Software & Tech. | | 0.03 | 0.04* | 0.09*** | 0.09*** | RE |
| | | (0.02) | (0.02) | (0.02) | (0.02) | |
| Writing | | 0.01 | 0.02 | 0.04* | 0.05** | RE |
| | | (0.02) | (0.02) | (0.02) | (0.02) | |
| **Complementarity** | | | | | | |
| Number of Complements | | | 0.06*** | 0.08*** | 0.08*** | 0.08*** |
| *Degree Centrality (log)* | | | (0.01) | (0.02) | (0.02) | (0.03) |
| Diversity of Complements | | | | 0.21*** | 0.22*** | 0.28*** |
| *1-Gini* | | | | (0.05) | (0.05) | (0.05) |
| Value of Complements | | | | | 0.76*** | 0.70*** |
| *Adjusted Neighbour Premium* | | | | | (0.04) | (0.05) |
| Constant | 1.23*** | 1.18*** | 1.03*** | 0.96*** | 0.71*** | 0.71*** |
| | (0.03) | (0.03) | (0.05) | (0.05) | (0.06) | (0.06) |
| Observations | 962 | 962 | 962 | 962 | 962 | 962 |
| Adjusted $R^2$ | 0.05 | 0.10 | 0.21 | 0.39 | 0.46 | 0.49 |
| *Note:* | | | | * p<0.1, ** p<0.05, *** p<0.01 | | |

Model 1 and 2 show the effect of supply and demand on the value of skills. Here we see, in alignment with our first hypothesis, that skills with lower supply and higher demand have higher values on average. In model 2 we add control dummies for the seven skill communities, while excluding the domain Admin as a reference group. We see that the value of a skill clearly depends on the community it belongs to, as all skills apart from Writing work have a higher price than our reference group, with Finance & Legal as the community of highest skill value on average. Accordingly, skills from Admin & Management work have the lowest values on average. In model 3 we show that the value of a skill is positively related to its number of complements. Model 4 brings diversity into the picture. Skills with a more diverse set of complements, tend to be of higher value. Model 5 introduces the average value of complements, which has a significant positive relation with skill values. Lastly, in model 5b we run a panel model controlling for differences across skill communities via random effects.[11] The complementary value of a skill plays a crucial role for its own value, as proposed in our second hypothesis. Interpreting the coefficients, we see that supply and demand still have a significant influence. On average, a one percent increase in the number of workers decreases the value of a skill by 27 percentage points while a one percent increase in demand elevates the value of a skill by 20 percentage points. Overall, we see that all hypothesised aspects – supply, demand, and complementarity – influence a skill's value. However, the changing coefficients of skill domains, when introducing the complementarity metric, indicate that the relevance of complementarity is differently distributed across skill domains, as shown in Figure 3, which plots the 962 skills' premia against the value of their complements.

---

[11] The findings in Figure 3B show that the relationship between the skill premium and the dependent variables is different across skill communities; see slope of scatter plots.

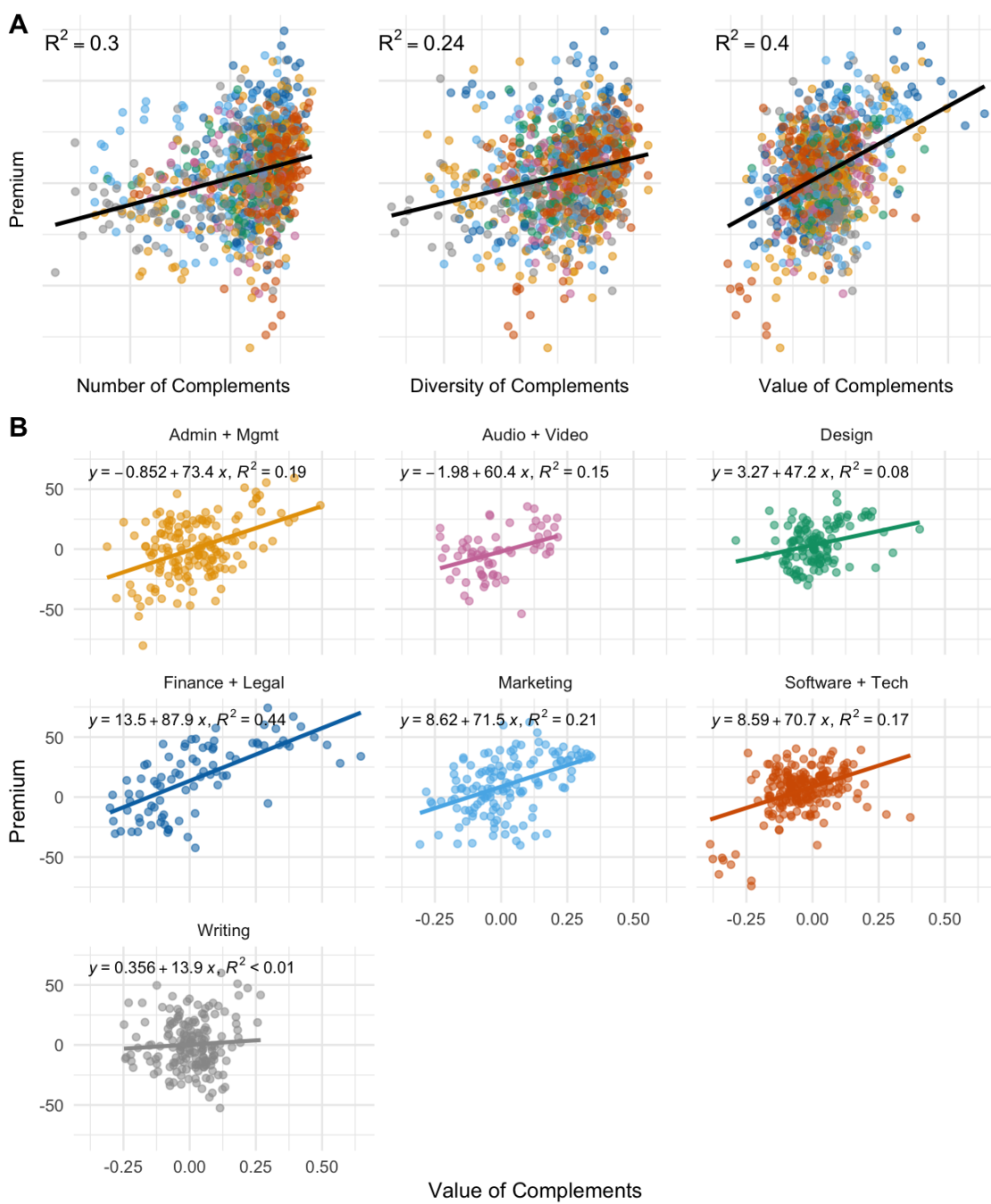

***Figure 3 (A)*** *The most valuable skills have high levels of complementarity, across all three dimensions; number, diversity, and value of complements. We clearly see that skills from specific domains, such as "Software & Tech" or "Finance & Legal", exhibit a significantly higher number, diversity and value of complements and, hence, higher skill values.* ***(B)*** *The relationship between a skill's value and the value of its complements is different across skill communities. For Finance & Legal and Marketing skills, a better complementarity translates much more strongly into higher skill values than in other skill communities.*

Here, we clearly see that the most valuable skills are skills with a strong complementarity; that is, those skills that are connected to a large and diverse set of skills of high value. We clearly observe that skills from specific communities, such as Software & Tech or Finance & Legal, exhibit significantly stronger complementarity and higher skill price. Accordingly, skills such as *Python* (Software and Tech) or *Business Coaching* (Legal and Finance) could be described as hubs in network terminology, as they can be combined with many other skills of high value.

At the same time, the results of our regression analysis confirm our assumption that – besides market forces of supply and demand – the value of a skill is influenced by the complementarity of the skill. However, our analysis also indicates that the benefits of strong skill complementarity are not distributed equally across skill communities, as illustrated in Figure 3B, which depicts the relationship between skills values and their complementarity for each of the seven skill communities. We see that both the baseline of skill values in each community (intercepts) and the benefit of a stronger complementarity (slope) are different across skill communities. We try to account for these differences across groups by introducing a random effects model in regression analysis 5b. In particular, the community of Finance & Legal and Marketing skills seems to be different to the other groups, as we see that the additional benefit of a stronger complementarity is significantly higher than in the other five communities. In Finance & Legal, furthermore, we observe a high initial level of skill values, which might indicate the strong signalling value of this community of skills, as they are highly formalised skills from a well-defined profession.

## *Understanding the Complementarity between Skills and Workers*

Our analysis indicates that the degree of complementarity of a skill influences its value in the marketplace. The higher the potential of a skill to be combined with many and diverse high-value complements, the higher its own value. While one can explain parts of the value of a skill in absolute terms, regardless of who applies it, specific combinations of skills have specific benefits as they work in a synergistic relationship (Anderson, 2017). In practice, skills are never applied in isolation. Workers add new skills to their existing skill portfolio, combining them into individual bundles of skills. In this scenario, the literature suggests that the value of skills (e.g. Anderson, 2017 or Neffke, 2019) depends on how they are combined – implying that a particular skill will not hold the same value for all workers. This builds on the intuitive idea that there are larger (or smaller) synergies between skills, depending on which skills are being combined. For example, compared to the average worker, a person who is experienced in *Logo Design* with *Adobe*

*Photoshop Pro*[12] might benefit disproportionately when learning about *Brand Strategy*, as this skill has strong complementarities with her existing skill set – in fact *Logo Design* is often part of a *Brand Strategy*. Similarly, there are dependencies between skills. *Python* is required to programme *Natural Language Processing* or *Adobe Photoshop Pro* can be used effectively for *Logo Design*. The skill network that we created already captures the result of these beneficial complementarities in action, as it shows us which skills are most frequently combined with each other. One could use the structure of the network as a piece of evidence to suggest that there is no single best, "most valuable", skill for everyone to learn, as we observe local hubs and clusters of skills rather than a centralised structure in which the entire network evolves around one "superstar" skill or group of very central and valuable skills.

In contrast to the *absolute* value of a skill, described in the previous section by a skill's demand, supply, and complementarity, here we elaborate on its *relative* values, based on the assumption that the value of a particular skill for a particular worker depends on its complements. To illustrate this relative value, we examine skill–worker combinations and the value of a particular skill in different such combinations. We have grouped the workers in our dataset, just like skills, into seven domains, which share the same labels as the skill communities. Workers are assigned to the domain with which they share the largest number of skills. (This quite clearly distinguishes workers, as for 78% percent of them at least half of their skills stem from one specific community.) We then measure the premium of each of our 962 skills depending on the domain of the worker who is applying them, to investigate beneficial complementarities across skill domains.[13] The findings of this analysis are summarised in Figure 4.

---

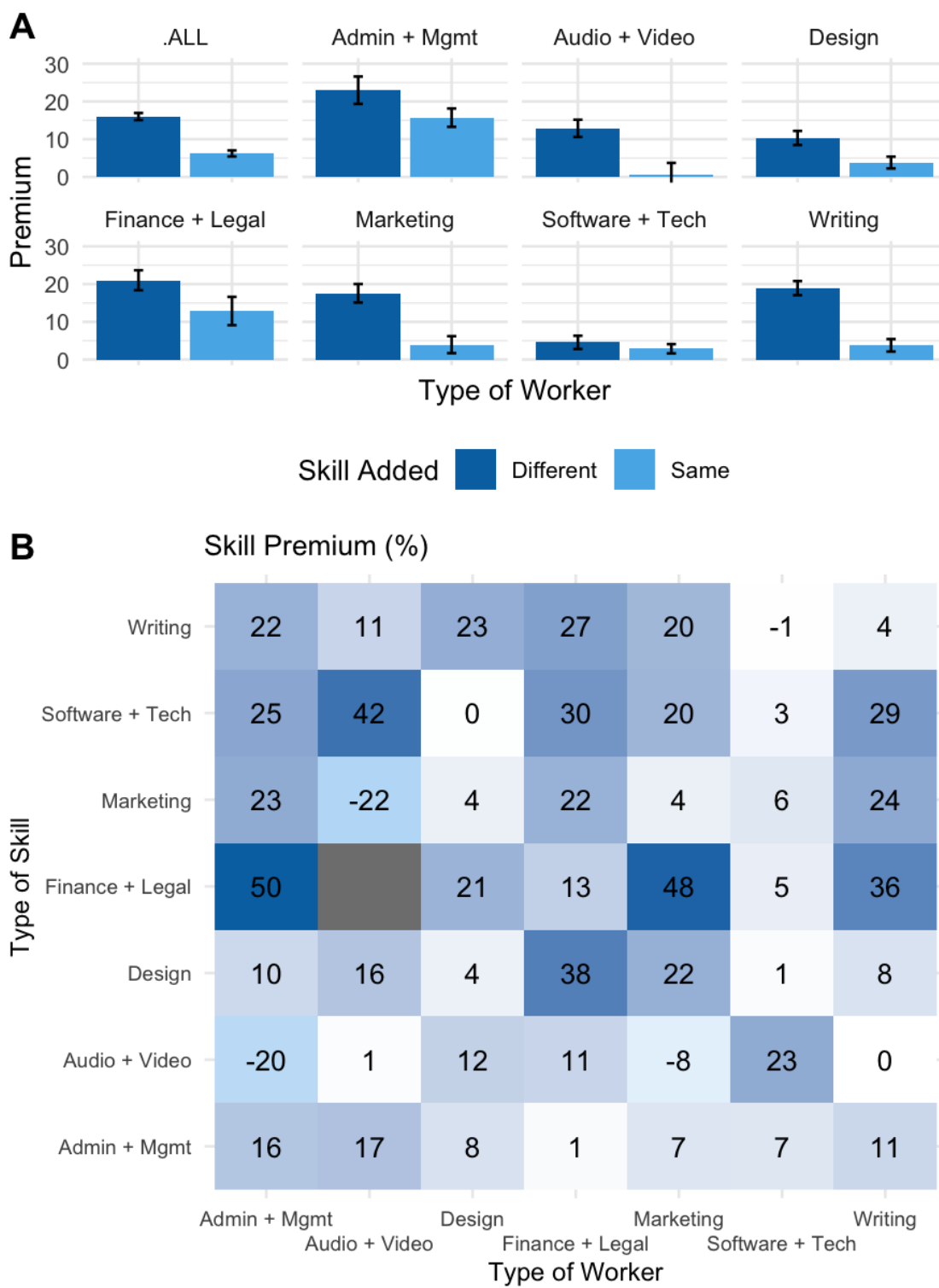

***Figure 4 (A)*** *On average, it pays off to have a diverse skill set, as the top left chart shows. However, skill diversity is more profitable for some workers (e.g. those in Writing and Marketing) than for others, like those in Finance & Legal or Software and Tech work. **(B)** Skill Premia: Here we see that relative to specialisation – the diagonal of the matrix – generalisation is profitable across skill and worker domains. But we clearly see that learning a skill from Software & Tech or Finance & Legal is generally more profitable for all workers.*

The general finding is that skills are more valuable when combined with skills of another community, as shown in Figure 4A. Likewise, the diagonal of the matrix Figure 4B shows the premium of skills from all seven communities, when applied by workers who predominantly have skills from the same domain. For some skills this effect is stronger than for others. For example, skills from the community of Software & Tech or Legal & Finance, are particularly profitable throughout all worker domains. In comparison to tech workers, Software & Tech skills are seven times more valuable for workers in Marketing and ten times as valuable for workers in Finance & Legal. This could indicate the high general purpose value of many skills from Software & Tech, in particular programming languages, which are beneficial in various combinations outside their original domain.

## *"Racing with the Machines": The Value of AI Skills*

At times of increasingly capable "machines", the value of skills is put at question. The fundamental transition unfolding on the labour market due to the widespread application of AI technologies is one of the central motivations for our work presented here. In the following, we illustrate the relevance of our work with the examples of AI skills. AI is widely considered to be a major breakthrough technology that is transforming the economy and society (OECD, 2021a). Hence, we assume that the skills that are required to work with AI technologies should be of high market value. We hand-labelled 42 skills[14] as AI skills based on the keyword list of AI-related terms developed by Righi et al. (2020). These skills either describe an AI technology, such as *Machine Learning*, or a prerequisite for working with an AI technology, such as *Python*. Of the 42 skills we labelled, only 21 AI skills remain in our analysis, as many of them only occur in a few projects and were therefore discounted (see Table A3 in the Appendix).

As complements to breakthrough technologies, AI skills are thought to benefit from the greater and more widespread application of AI applications, such as large language models or image generation (E. W. Felten et al., 2019; Squicciarini & Nachtigall, 2021).

---

[14] These skills are: 'advanced-analytics', 'ai', 'algorithm-development', 'algorithms', 'analytics', 'apache-spark', 'artificial-intelligence', 'artificial-neural-networks', 'automation', 'automation-software-release', 'big-data', 'bot-development', 'c++', 'chatbot-development', 'cloud-computing', 'clustering', 'computer-vision', 'data-analysis', 'data-analytics', 'data-engineering', 'data-science', 'database-architecture', 'deep-learning', 'deep-neural-networks', 'ibm-watson', 'image-processing', 'imageobject-recognition', 'java', 'keras', 'machine-learning', 'machine-learning-model', 'natural-language-processing', 'natural-language-toolkit-nltk', 'neural-networks', 'pattern-recognition', 'python', 'python-script', 'robotic-process-automation', 'robotics', 'supervised-learning', 'tensorflow'.

Recently, these applications have begun to enter more and more domains of knowledge work, such as creative writing and illustration design, as well as legal work or financial reporting. Accordingly, we postulate that AI skills should have strong complementarities, that is they should be combinable with a flexible and valuable set of adjacent skills. This complementarity is a central aspect in discussions about the sustainability of work. Ideally, newly learned skills should have the feature of complementing AI technologies, allowing workers to "run with the machines" (Brynjolfsson & McAfee, 2012), rather than being subject to automation (more on the comparison between AI exposure and economic value can be found in the Appendix S3). We assume that strong complementarity and a sustained level of demand make AI skills extremely valuable.

With the use of the metrics developed in this work, we examine this assumption and summarise the findings in Figure 5. First, we compare the skill premium for AI skills with the average skill premium in our dataset. Similarly, we examine aspects of demand and supply, as well as our three complementarity metrics, for the example of 20 AI skills. The rich granularity of our data further allows us to examine the value of individual AI skills. Lastly, as technological advancements unfold rapidly, we illustrate the development of skill values over time. We use the example of AI skills to illustrate that the price of skills is not stable over time, as this is influenced by changes in supply and demand.

The findings presented in Figure 5A show that AI skills are special. With a premium of 21%, AI skills are far more valuable than the average skill in our sample (8%). Reflecting on our previous findings, we speculate that some of the AI premium can be explained by complementarities. Indeed, AI skills have an above average number of complements of large diversity. Perhaps most importantly, the complements are significantly more valuable than usual, regarding our data. In addition, part of the AI premium stems from a sustained demand, as our data indicates that AI skills have a significantly higher level of demand over supply than the average skill in our data set. In general, AI skills are clearly among the most valuable of all skills in our dataset. However, our analysis also reveals sizable differences in the value, once we examine AI skills separately (B). We see that skills around machine learning – ML (40%), Tensor Flow (38%), Deep Learning (27%), NLP (19%) – are more valuable than skills around Data Analysis (14%) and Data Science (17%), followed by the most prominent programming languages used to build AI, like Python (8%), C++ (7%) or Java (5%).

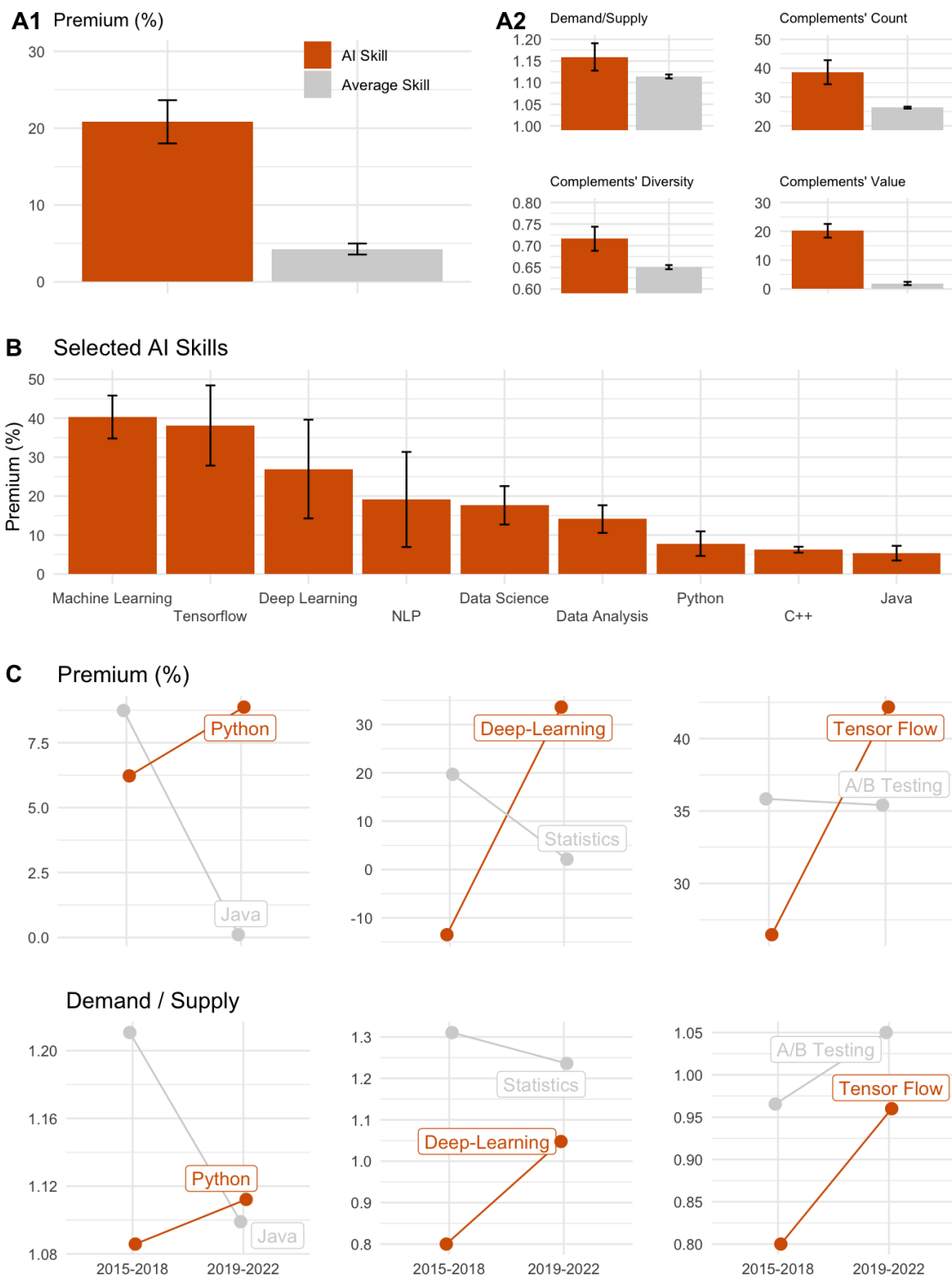

*Figure 5* **(A1)** *Working with AI pays off: The value of AI skills is significantly higher than for average skills.* **(A2)** *This is likely due to the high level of relative demand over supply and the strong complementarity of AI skills.* **(B)** *When examining the value of AI skills individually, we see that skills around machine learning – that is, ML (40%), Tensor Flow (38%), Deep Learning (27%), NLP (19%) – are more valuable than skills around Data Analysis (14%) and Data Science (17%), followed by the most prominent programming languages used to build AI, like Python (8%), C++ (7%) and Java (5%).* **(C)** *Over time, we observe that*

*changes in the value of a skill are closely aligned with changing supply and demand. As demand (relative to supply) for Python, Deep Learning, and TensorFlow increased (lower panel), the value for these skills increased (upper panel).*

Finally, the bottom panel of Figure 5 illustrates this change by comparing the premium of selected skills between the periods 2015–2018 and 2019–2022 (average premium for each period). First, we contrast the change in skill premium for two popular programming languages, *Python* and *Java*. While both languages start with more or less the same premium in the timespan of 2015–2019, they develop differently over time. In our data set, *Java* significantly decreases in premium, *Python* gains in market value. A driver of this trend could be a change in popularity, as *Python* has been rising to become *the* data science super skill over the last decade (Grus, 2019). This has not been the case for *Java*, as the comparison of demand (number of projects) versus supply (number of workers) indicates in the lower panel. A similar logic applies to AI skills in the field of deep learning. In panels B and C, we show the changes in premium of *Deep Learning* and *TensorFlow*, one of the most commonly used software environments in which to program deep learning applications. As a comparison, we selected the skill *Statistics* and *A/B Testing*, one of the most frequently used conventional applications of statistics. In our freelancer sample, we see how the premium for *Deep Learning* and *TensorFlow* increased significantly over the last 10 years, while the value of *Statistics* and *A/B Testing* stagnated. Similarly to the programming languages, our online gig workers sample shows a significant rise in the demand relative to supply for *Deep Learning* and *TensorFlow* compared to the skills *Statistics* and *A/B Testing*. This analysis illustrates that the relationship between market demand and supply and skill premia holds true over time. This finding is based on the number of workers and projects in our dataset and should not easily be generalised to the wider economy.

# Discussion

## Conclusion

Technological change does not affect all tasks and occupations equally. New technologies require new skills while making others redundant. In other words, technological change is not "skill-neutral". As a result, the skill composition of occupations changes. To respond effectively, workers need to reskill, however, the precise skill requirements of newly emerging jobs and the economic benefits of learning a new skill are often uncertain and constantly evolving. It is therefore difficult for workers, employers, and policy-makers to build profitable and sustainable reskilling pathways. To address this uncertainty, we propose a method that attaches a market value to skills based on market demand and supply as well as their complementarity with other skills.

This expands the understanding of human capital formation in at least two ways. First, we reveal the market value of individual skills to then investigate what drives the variation in skill values. We find that the value of a skill depends (obviously) on market forces of supply and demand, but we find that it is also strongly correlated with a skill's complements. Skills tend to be valuable if they are frequently combined with a diverse set of other valuable skills. According to our analysis, such skills include programming languages like *Python* or generalistic legal skills like *Contract Law*. In addition to the absolute value of a skill, we reveal a relative value that emerges from beneficial combinations of skills. Often skills are most valuable if combined with skills from a different domain. For a set of AI skills we illustrate how to apply our evaluation metric. We show that skills needed to construct and maintain AI, which is widely considered to be a major breakthrough technology, have significantly higher skill values than the other skills in our dataset—With a premium of 21%, AI skills are far more valuable than the average skill in our sample (8%). AI skills have an above average number of complements of large diversity, since AI technologies enter more and more domains for knowledge work. Furthermore, we track the development of skill values over time and find that AI skills, such as *Deep Learning* and *Python* have been gaining in value significantly in recent years. Our model allows us to ascribe these changes to an increase in demand relative to supply.

This study has certain limitations. Firstly, our investigation provides no real identification strategy. We therefore try to avoid making causal statements throughout the text. Our findings only indicate an association between complementarity and skill values, as there is no source of exogenous variation on complementarity. Secondly, we measure the market value of skills *relative* to each other in a specific setting. This allows for comparisons between skills, but statements on the absolute value of skills are not possible, and in any case would not be easily transferred from the particular context of our dataset to a different one. That said, while a statement such as "acquiring deep-learning skills will increase my wage by 27%" (see Figure 5) would be problematic, stating that "deep-learning might be a more valuable skill to learn than data-analytics" is in line with our approach. The third limitation relates to explanations. In this paper we find that the market value of a skill is relative, as it depends on how it is combined with other skills. While this makes intuitive sense and is highly relevant for reskilling purposes, it is outside the scope of this paper to develop a theoretical framework and empirical strategy to explain the how and why of these mechanisms. Finally, this work is based on data from online freelancing which comes with several advantages but also a number of pitfalls. Most notably, online freelancing is limited to fully digital professional services, meaning that many (predominantly manual) occupations are missing entirely. Also, online freelancers are solo workers. They do not have organisational roles or interactions with co-workers, which certainly add relevant dimensions to skill complementarities, in particular between cognitive, non-cognitive and social skills. The discussion on complementarities between skills alone is useful for the types of occupation covered in this work, such as short-term IT, creative or translation gigs. However, in other settings, e.g., R&D or product design, cross-worker complementarities emerge from the interaction of a set of specialised workers, as shown by Neffke (2019). Both perspectives could be combined in future investigations. At this point it is worth noticing that in the conventional economy artificial restrictions on either supply or demand might apply as, for example, lawyers or doctors require official government certification. These restrictions do not play out in the observed online freelance market, which should be taken into consideration when trying to generalise our findings. Moreover, relative to the size of the overall workforce, rather few people work via online freelancing platforms. As outlined in the Method section, we argue that online freelancing data is certainly context specific, but the fundamental dynamics of demand, supply, and complementarity that drive the value of skills should not be drastically different in other segments of the labour market. That said, with adequate data access, our methodology could be extended to other data sources, such as online job advertisements and online career platforms, which cover large parts of the conventional labour market.

## Policy Implications

Our work on categorising and evaluating skills allows for multiple advances in understanding labour market developments. It can help to establish a taxonomy of skills, understand their application and individual complementarity, and enable automated, individual, and far-sighted suggestions on the value of learning a new skill in a future of technological disruption. Hence, various potential policy implications and applications could be imagined.

First, reskilling institutions, like the European Centre for the Development of Vocational Training, could be the main beneficiaries of this highly individualised data. The high granularity of online generated data allows us to describe the skill profiles of individual workers and track their development over time. It also enables reskilling institutions to assess the individual complementarities of learning a new skill. Building on online generated labour market data, workers with a need to reskill could insert their current skill profile, be located in the landscape of skills, and receive targeted reskilling advice. This would allow them to switch to more sustainable occupations that are closely related to their existing skill set with minimal reskilling effort. Via these individualised reskilling recommendations education providers and vocational training organisations could address the urgent need for *individualised* solutions in adult reskilling. Furthermore, the continuous "pricing" of skills over time enabled by our approach allows reskilling practitioners to monitor the development of skill values and advise workers on which reskilling to "invest" in.

Secondly, official occupational and skill taxonomies could be improved with near real-time online generated data. As technology creates a demand for novel skills, new occupational clusters can quickly emerge and pull away from official taxonomies, such as the European Skills, Competences, and Occupations (ESCO). This will be bad news for both firms and workers, if professional training providers find it hard to "speak" in the same language as market demand. Online generated data, by contrast, stems from most recent market developments and allows for identification of new occupational clusters, including in-demand skills. These data-driven, near-real time taxonomies could complement conventional classifications. An immediate contribution to current policy efforts would be the continuous (re-)classification of AI and "green" skills or jobs, as the "twin-transition" has been identified as a catalyst for active labour market policies (OECD, 2021b).

*Outlook*

This work uses data from online labour markets. The major advantage of this data source is the availability of granular information on demand, supply, and wages – and hence on the value of skills. On the other hand, this data covers only one segment of the labour market. That said, the methodology presented here could easily be adapted to analyse other data sources covering larger parts of the labour market, such as job vacancy platforms (e.g. Indeed) and social career platforms (e.g. LinkedIn). Our results have implications for the debate on successful reskilling strategies in times of dynamic technological change. We find that in-demand skills pay off, and that they can be identified. This suggests that individualised skill-centred reskilling pathways could represent a promising avenue to mitigate skill mismatches – assuming access to timely and granular data.

The European Commission has recognised the need and potential of a data-driven approach to closing the skill gap by bringing forward various legislative and policy proposals. The Pact for Skills (European Commission, 2020), launched in 2020, for example, aims to maximise the impact and effectiveness of skills investment, with a particular focus on upskilling and reskilling in the vocational training sector. For a successful implementation of the Pact, two aspects are crucial. Firstly, industry needs for specific skills must be made explicit, and secondly, the unique training histories of individual workers need to be acknowledged. Online generated data of worker profiles presents a promising approach to monitoring occupation taxonomies and skill requirements via online labour platform data. This is very much aligned with Europe's interest in further building out their skill foresight (French Presidency of the Council of the European Union, 2022) via skill anticipation and support for career transitions. It can offer targeted and near-real time reskilling advice to workers, regarding both industry needs and the worker skills required to fulfil them. This action could support the Commission's proposals (European Commission, 2021) for recommendations on individual learning accounts and micro-credentials, fostering the skill-by-skill learning of workers instead of traditional certification.

Similarly, the Commission's 2022 Data Act (European Commission, 2022) has identified the importance as well as the complications of accessing business (and platform) data in the interests of the public, while acknowledging the protection of business interests.

However, the retrieval and usage of private sector data, such as online labour market or job vacancy data, by public body institutions, is not necessarily enabled under the new legislation. Enforced sharing of private sector data requires the ex-ante proof of a "public emergency", and the current modes of automated data retrieval, such as web-scraping, could even be prohibited by the Data Act if they were to be interpreted as coercive or deceptive according to Article 11. In light of this well-intended but potentially contradictory proposal to current EU data legislation, future amendments need to land on a mechanism that gives public bodies acting in the interest of the public the right to access data.

Our investigations on the complex ecosystem of skill formation show that online data can be a valuable tool for designing sustainable reskilling policies. To leverage the full potential of this resource future legislation needs to make public interest its focal point, allowing data access via web-scraping or research APIs, while enabling strategic public–private partnerships to release the full potential of online generated data for the benefit of society.

## CRediT authorship contribution statement

**Fabian Stephany**: Conceptualization, Methodology, Software, Validation, Formal Analysis, Investigation, Data Curation, Writing – Original Draft, Writing – Review & Editing, Visualisation, Supervision, Project Administration, Funding Acquisition. **Ole Teutloff**: Conceptualization, Methodology, Software, Formal Analysis, Investigation, Data Curation, Writing – Original Draft, Writing – Review & Editing.

## *Acknowledgements*

The authors are very thankful to Estrella Gómez Herrera for an extensive collegial review of the work. Similarly, we thank Ingo Zettler and Magnus Lindgaard Nielsen for their feedback on a presentation of the paper and to Hendrik Send and Georg von Richthofen for initial brainstorming on the work. The authors are grateful for the wonderful copy editing work by David Sutcliffe. The authors furthermore are grateful to Fabian Braesemann for curation of the underlying data, which has been retrieved under the John Fell Oxford University Press Research Fund, grant number 0008391. Fabian Stephany thanks the ESRC Digit Innovation Fund (G2781-15) for the financial support of his work.

# Appendix

## *S1: About Online Labour Market Data*

With the rise of digital platforms as economic intermediaries more and more data about labour market processes has become available online. Indeed, as automated retrieval of large amounts of this economic transaction data has become more feasible in recent years, the study of labour markets with digital trace data is gaining momentum. Besides relying on traditional survey data, researchers have thus started to investigate skill developments with these new digital data sources as well.

While the idea proposed here of using online labour market data for skill monitoring is novel, a number of scholars have explored other sources of online generated data to investigate skill formation. De Mauro et al. (2018), for example, examined the skill complexity of the new profession of data science with data retrieved from various job boards. Similarly, both Börner et al. (2018) and Calanca et al. (2019) demonstrate the increasing relevance of soft skills based on online job vacancy data. Bastian et al. (2014), on the other hand, make use of data from LinkedIn to compare the relevance of certain "hard skills" across industry domains. While the methodology we propose in this paper to study skill development and detect the emergence of novel occupational domains is applicable beyond online labour market data, to things like data from online job vacancy portals (such as Indeed or Glassdoor) or data from professional social network sites (like LinkedIn), these different data sources have particular advantages and shortcomings, as summarised in Table A1.

**Table A1.** Suitability of online data sources to the study of the demand and supply sides of work. Compared to data from online job vacancy sites and career portals, online labour market data allow for the study of both the demand and supply side of work, including relevant information on prices. Source: Stephany & Luckin (2022).

| Data Source | Demand Side | Supply Side | Price Information | Broad Coverage |
|---|---|---|---|---|
| *Online Job Vacancies* | 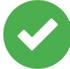 | 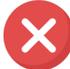 | 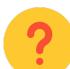 | 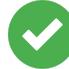 |
| *Networking Sites* | 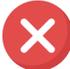 | 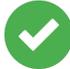 | 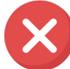 | 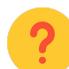 |

| Online Labour Platforms | 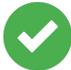 | 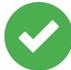 | 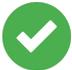 | 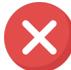 |
|---|---|---|---|---|

While online job vacancies cover a large segment of the labour market, including many industry sectors and also potentially non-digital and manual work, they seldom include information on price levels and give no indication of the possible supply in the targeted population. Data from professional social media sites, like LinkedIn, on the other hand, allow for an in-depth analysis of skill compositions in the population. However, no price or income information is revealed, and matching efficiencies cannot be evaluated in the absence of demand-side data. Online labour market data, e.g. from platforms like UpWork or Fiverr, only covers a small segment of the labour market, namely digitised tasks from jobs in the professional service sector. However, these data have the major advantage of containing information on both the demand and supply side of skills. In addition, it is possible to observe the matching process and price, e.g., hourly rate, for each job with a particular skill bundle attached to it.

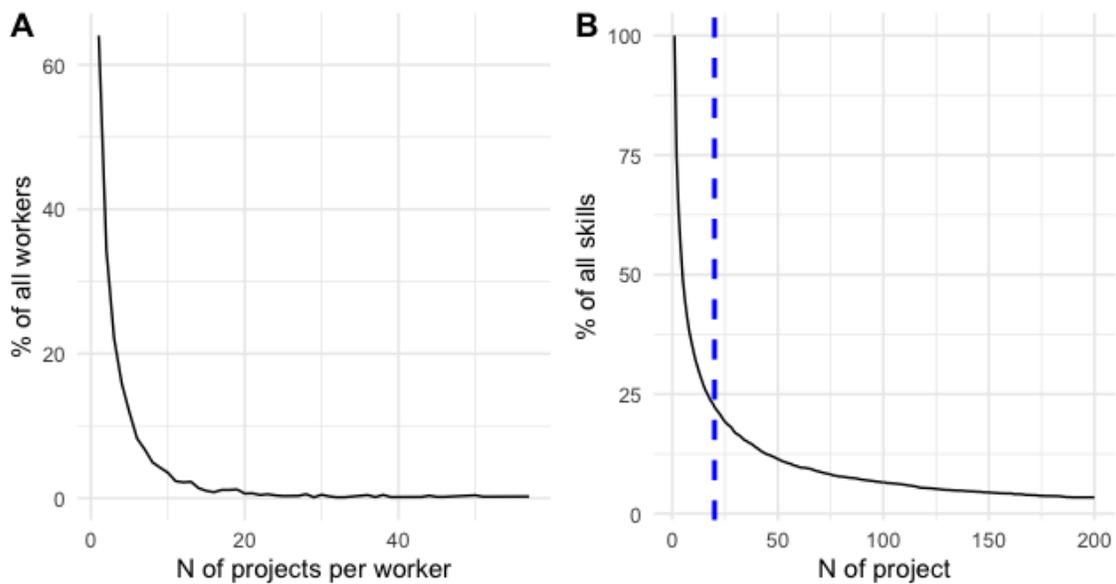

***Figure A1 (A)*** *The minority of workers (4%) has had more than 10 projects and more than half of the workers had only two projects or less. **(B)** For skills, we see that the largest share of skills only occurred in a few projects; only 22% of all skills were applied to 20 projects or more.*

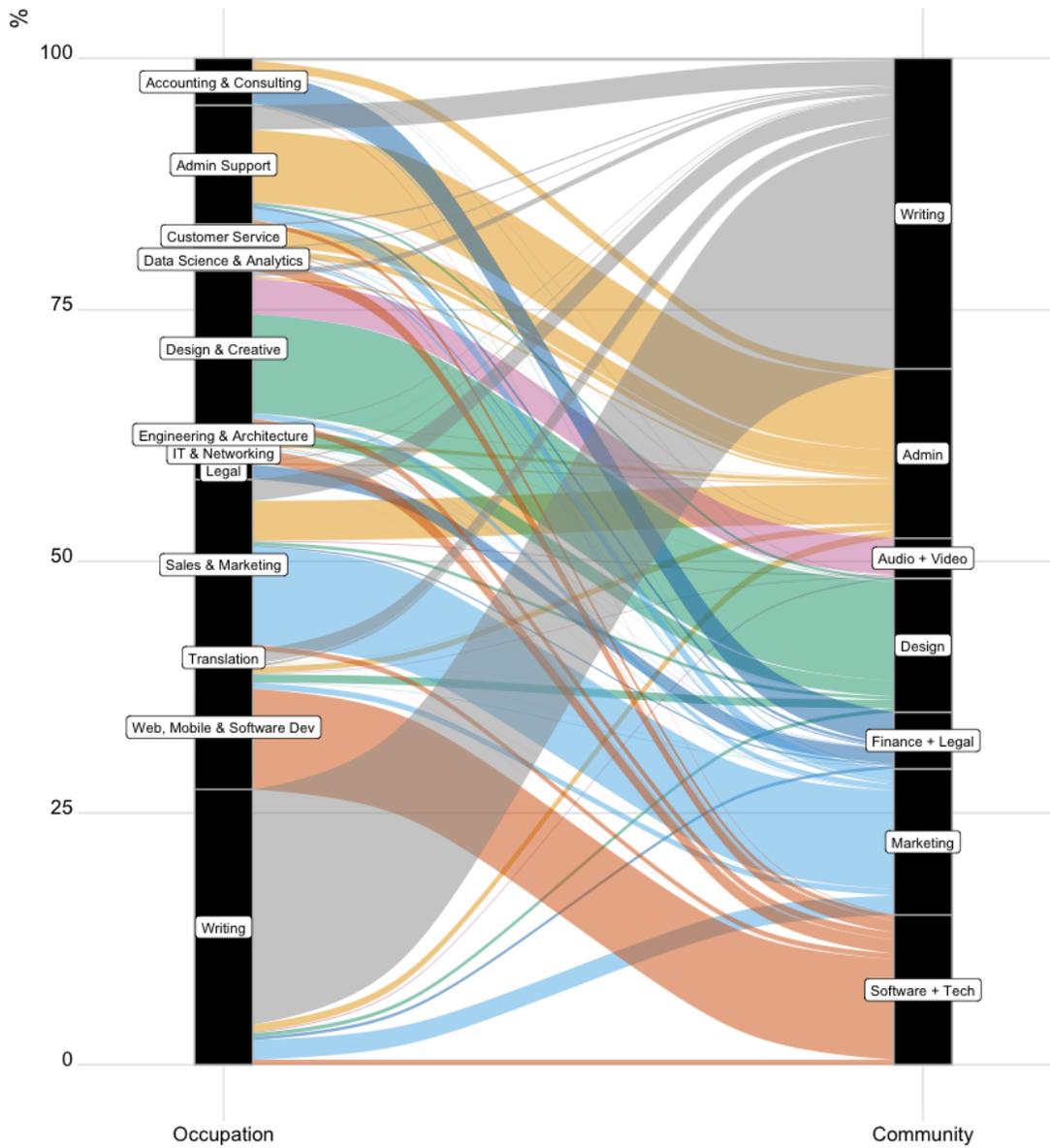

**Figure A2** *Skill application clusters largely overlap with the occupation taxonomy provided by the freelance platform, i.e., the majority of skills in the application cluster of "Design" also fall into the occupation category "Design & Creative".*

## S2: Calculating the Price of a Skill

The "ratio" of each skill is calculated by comparing the mean rate of projects that require the respective skill with the mean rate of those projects that do not require it:

$$ratio = \frac{\sum_{i=1}^{n} wage/n}{\sum_{j=1}^{m} wage/m} - 1$$

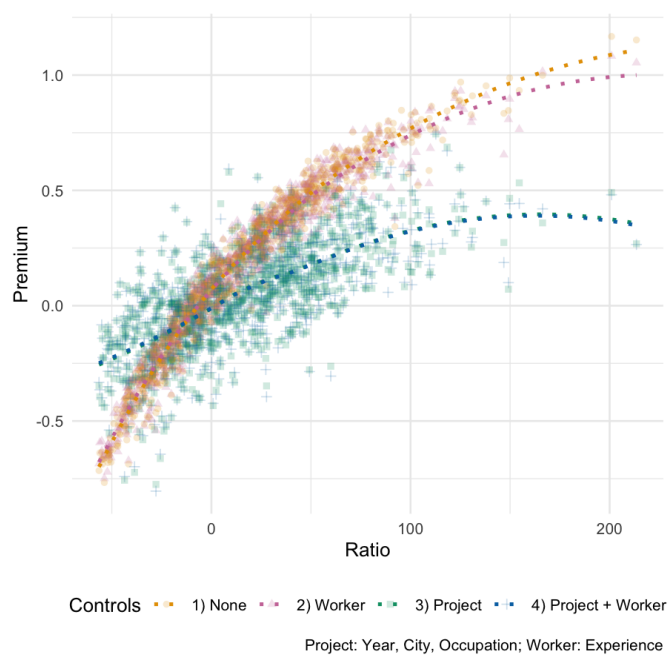

***Figure A3*** *In this robustness check, we compare the wage ratio with the premium of skills. We perform this test for various specifications of the wage regression. When applying no controls, premium and ratio are strongly correlated (0.98). Controlling for the worker's experience slightly diminishes the correlation (0.97). It is the controls of location, year, and occupation that adjust the premium significantly relatively to the ratio (0.70)*

Based on the theory of skill complementarity (Anderson, 2017; Stephany, 2021) the systemic value of a skill should depend on the value of its closest complements, that is, the other skills that it is frequently combined with. The idea is that some of the variation in the premium of skill A should be explainable by the variation in premium of the most adjacent set of skills (B), see Figure A4.

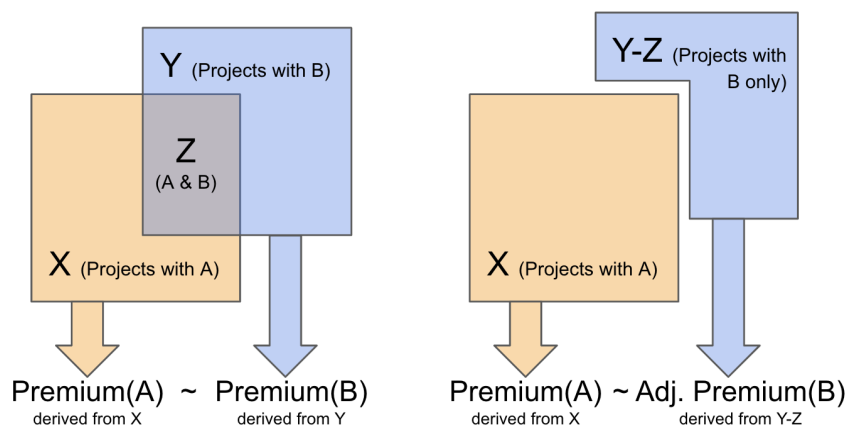

**Figure A4.** *LHS: The premium of skill A and B might correlate to some extent, as both skills are jointly applied in the same set of projects (Z). RHS: To exclude the effect that the joint application has on the correlation of skill premium, we propose an adjustment that takes only the non-overlapping project set into account (Y–Z).*

Given the construction of our premium metric, some of the correlation in skill premia might be explained by the fact that both skill A and skill B stem from a joint set of projects (Z) with the same wage (Figure A4, left hand side). However, ideally, we would like the premia of B to be determined only based on the variation of projects wages from set Y–Z. To achieve this property, we propose a two-stage regression solution. First, we estimate the premium of skill A and B, as described in the section above. Then, we estimate the premium of B with the average project wage level of X. The resulting regression residual Adj. Premium (B) should be free of any variation that is attributable to the overlapping project set X and should only be explainable by the variation from the project set Y. We call this residual value the adjusted complements' premium. To test the effectiveness of our adjustment, we correlated the skill premium for all skills and a set of complements of different sizes (1-5 complements). We performed this procedure for both the adjusted and the unadjusted complements' premium (for more details see Figure A5). The resulting test also reveals that the optimal number of complements to consider in terms of explanatory power is three.

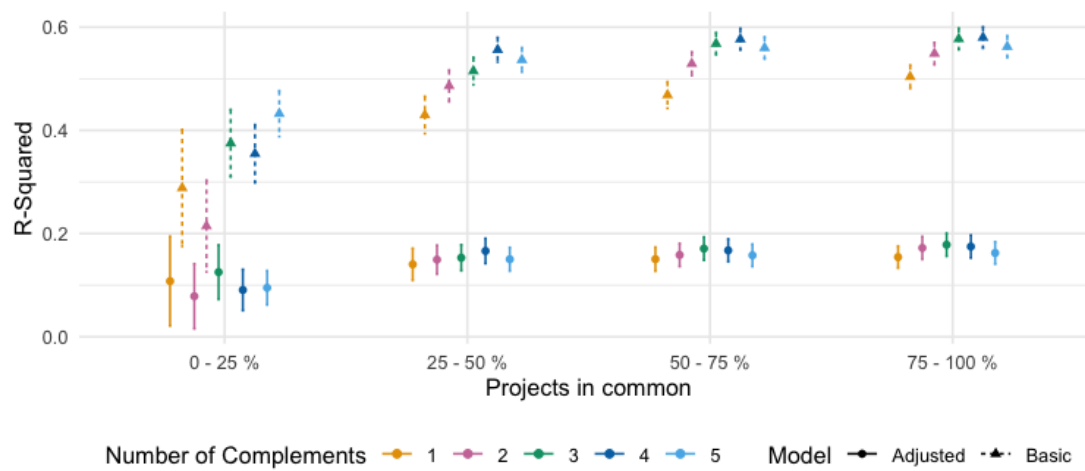

***Figure A5*** *This chart shows how much the premium of a skill can be explained by the premium of its complements (R-Squared). In the upper half of the chart, we see that the explanatory power increases – from left (0–25%) to right (75–100%) – with the share of projects that both skill and complements have in common. To control for this, we calculate the adjusted premium of complements, which does not include the information from the overlapping projects (lower half). While the explanatory power diminishes, we see that it no longer reacts to the share of overlapping projects.*

***Table A2*** *The top and bottom 20 skills are ranked by their premium. Many of the top ranked skills stem from Finance & Legal while bottom ranked skills are from Admin work.*

| Rank | Skill | Ratio (%) | Price (USD/h) |
|---|---|---|---|
| 1 | business-coaching | 197.59 | 1.62 |
| 2 | patent-law | 163.35 | 1.52 |
| 3 | international-tax-law | 151.77 | 1.56 |
| 4 | corporate-law | 146.99 | 1.55 |
| 5 | non-disclosure-agreements | 146.41 | 1.11 |
| 6 | international-law | 144.14 | 1.21 |
| 7 | financial-projection | 135.29 | 1.58 |
| 8 | contract-law | 128.49 | 1.34 |
| 9 | trademark-consulting | 127.98 | 1.32 |
| 10 | financial-forecasts | 124.19 | 1.54 |
| 11 | financial-modelling | 122.69 | 1.62 |
| 12 | branding-strategy | 122.23 | 1.82 |
| 13 | cloud-computing | 121.75 | 1.44 |
| 14 | docker | 119.86 | 1.39 |

| 15 | financial-plans | 119.72 | 1.56 |
| 16 | venture-capital-consulting | 115.95 | 1.96 |
| 17 | deep-learning | 113.73 | 1.31 |
| 18 | business-coaching | 110.33 | 2.10 |
| 19 | employment-law | 108.16 | 1.31 |
| 20 | legal-consulting | 107.64 | 1.14 |
| ... | ... | ... | ... |
| 943 | teaching-english | -49.60 | 0.75 |
| 944 | administrative-support | -51.10 | 0.87 |
| 945 | article-spinning | -51.32 | 0.78 |
| 946 | email-handling | -51.78 | 0.86 |
| 947 | video-upload | -51.87 | 0.58 |
| 948 | virtual-assistant | -51.99 | 0.84 |
| 949 | customer-service | -52.21 | 0.83 |
| 950 | chat-support | -52.38 | 0.91 |
| 951 | customer-support | -52.41 | 0.89 |
| 952 | translation-spanish-english | -52.48 | 0.75 |
| 953 | news-writing-style | -52.50 | 0.70 |
| 954 | active-listening | -52.78 | 0.83 |
| 955 | data-entry | -54.26 | 0.65 |
| 956 | call-handling | -54.71 | 0.79 |
| 957 | telephone-skills | -55.02 | 0.80 |
| 958 | phone-support | -55.14 | 0.87 |
| 959 | forum-posting | -55.51 | 0.67 |
| 960 | online-help | -55.58 | 0.92 |
| 961 | english-tutoring | -56.53 | 0.69 |
| 962 | order-processing | -56.92 | 0.72 |

**Table A3** *Of the 42 initially identified AI skills, 17 remain after filtering for a minimum of 20 occurrences. They are ordered by their premium.*

| Rank | Skill | Ratio (%) | Price (USD/h) |
|---|---|---|---|
| 1 | cloud-computing | 121.75 | 1.44 |

| 2 | deep-learning | 113.73 | 1.31 |
|---|---|---|---|
| 3 | tensorflow | 107.10 | 1.46 |
| 4 | natural-language-processing | 101.52 | 1.21 |
| 5 | machine-learning | 101.45 | 1.50 |
| 6 | database-architecture | 97.77 | 1.17 |
| 7 | artificial-neural-networks | 96.66 | 1.21 |
| 8 | artificial-intelligence | 74.57 | 1.17 |
| 9 | algorithms | 72.21 | 1.36 |
| 10 | data-science | 65.51 | 1.19 |
| 11 | automation | 51.01 | 1.36 |
| 12 | python | 48.84 | 1.08 |
| 13 | data-analysis | 39.91 | 1.15 |
| 14 | analytics | 38.18 | 1.28 |
| 15 | java | 28.94 | 1.05 |
| 16 | image-processing | 9.16 | 0.95 |
| 17 | c++ | 3.74 | 1.06 |

## S3: Calculating the AI Exposure of a Skill

Besides economic profitability, AI complementarity is certainly another relevant aspect under which the "future readiness" of a skill could be evaluated. A skill that is highly valuable but most likely to be automated by machines or algorithms in the near future might not be an optimal destination for reskilling interests. Moreover, a skill with low likelihood of automation but little economic reward might be similarly unappealing. For a comparison of our proposed skill evaluation metric, we calculate how much skills are exposed to AI technologies using the AI exposure values developed by Felten et al. (2021). In this work, the authors create an occupational-level measure of exposure to AI by linking 10 specific applications of AI identified by the Electronic Frontier Foundation (EFF) to O*NET workplace abilities. The authors then aggregate the resulting ability-level exposure scores to occupations at the six-digit SOC level. While this method identifies the relative exposure to AI, it remains agnostic about whether AI might substitute or complement a given ability. The method proposed by Felten et al. (2021)

therefore differs from previous approaches that focus on automation risk at the occupation-level (such as for example Frey & Osborne, 2017). We use the matching table developed by Braesemann et al. (2021) to map the six-digit SOC codes to the 98 platform occupations identified in our dataset. After the matching, we calculate the "AI exposure" for each skill via the following formula:

$$exposure\,(skill\,x) = \sum_{j=1}^{m} exposure\,(occupation_j)/n_j$$

Here the AI exposure of $skill\,x$ is given by the average of the AI exposure of all occupations $j = 1,...,m$ that contain $skill\,x$ weighted by the share of projects $n$ that are listed under the respective occupation. Our new metric allows us to compare both premium and exposure to AI for each skill, as illustrated in Figures A6 and A7. Our comparison shows that skills do not group randomly in the space of premium and AI exposure. Broadly speaking, we can see a tendency that premium and AI exposure are positively related, that is, that the most profitable skills tend to have a higher exposure to AI. Many skills from the domain of Software & Tech, including most of the AI skills, are of relatively high market value and show a high exposure to AI. Writing and Admin skills are of low market value and mid-level exposure to AI. Lastly, skills from Design or Audio & Video have low market value and low exposure to AI. Our group of AI skills are clearly different from the rest of all other skills, as they have above average premia and high exposure to AI. While the AI exposure score developed by Felten et al. (2021) does not allow for reliable statements on skill substitution or skill enhancement through AI, it is likely that at the current stage, most of the AI skills presented here benefit from high complementarity towards AI.

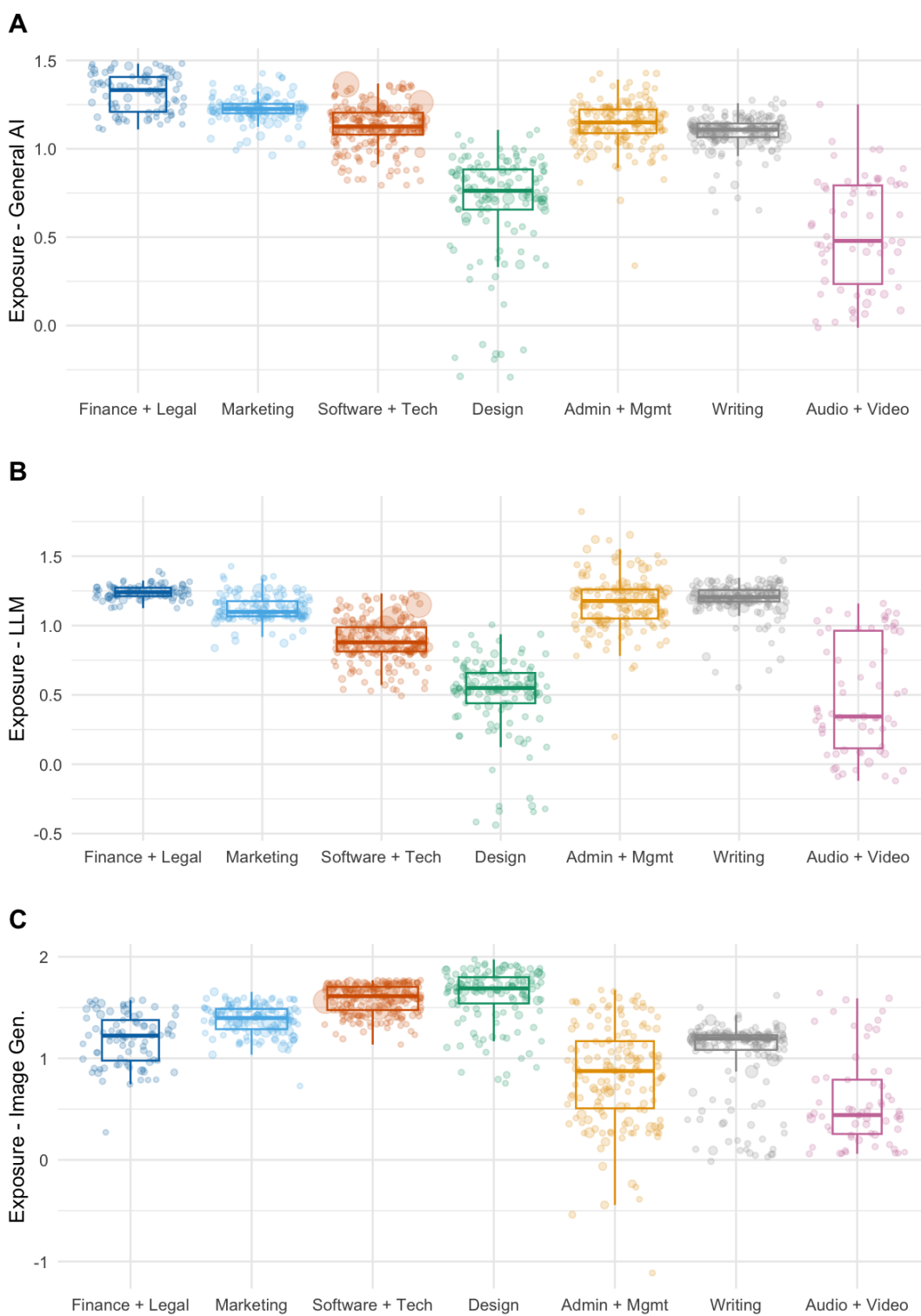

**Figure A6** *Exposure of skills to Artificial Intelligence by skill community. (A) The AI exposure of skills. (B) The exposure of skills to the AI subfield "Language Modeling". This includes large language models such as ChatGPT. (C) The exposure of skills to the AI subfield "Image generation". We can see that the exposure to different fields of AI differs significantly. For example, skills in the Design cluster have high exposure to image generation AI but rather low exposure to language modelling AI. The method for LLM exposure and Image generation exposure is based on Felten et al. (2023).*

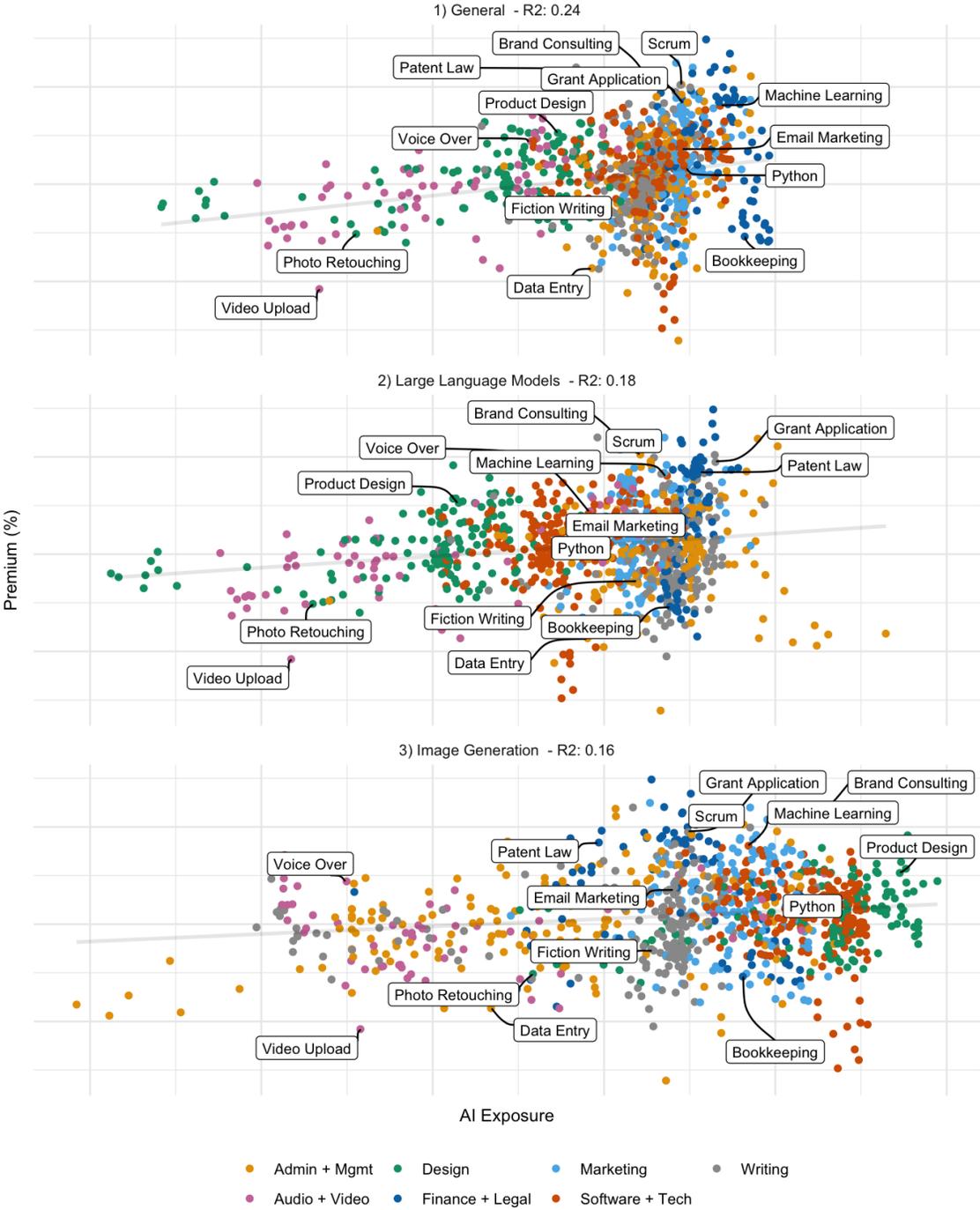

**Figure A7** *The premium and AI exposure of a skill are only weakly correlated. For the case of exposure to general AI models (1), we see a slight positive correlation with economic value: more valuable skills tend to have higher exposure levels to general purpose AI.*